\documentclass[%
 reprint,
 amsmath,amssymb,
 aps,
]{revtex4-2}

\usepackage{graphicx}
\usepackage{dcolumn}
\usepackage{bm}
\usepackage{lipsum}
\usepackage{xcolor}

\usepackage{float}
\usepackage{hyperref}
\hypersetup{
    colorlinks=true,
    linkcolor=blue,
    citecolor=magenta,      
    urlcolor=cyan,
    }

\begin{document}

\preprint{APS/123-QED}

\title{High-order aberrations of vortex constellations}

\author{R. F. Barros}
\affiliation{Tampere University, Photonics Laboratory, Physics Unit, Tampere, FI-33720, Finland}
 \email{rafael.barros@tuni.fi}
\author{S. Bej}%
\affiliation{Tampere University, Photonics Laboratory, Physics Unit, Tampere, FI-33720, Finland}
\author{M. Hiekkamäki}%
\affiliation{Tampere University, Photonics Laboratory, Physics Unit, Tampere, FI-33720, Finland}
\author{M. Ornigotti}%
\affiliation{Tampere University, Photonics Laboratory, Physics Unit, Tampere, FI-33720, Finland}
\author{R. Fickler}
\affiliation{Tampere University, Photonics Laboratory, Physics Unit, Tampere, FI-33720, Finland}

\date{\today}

\begin{abstract}

\noindent
When reflected from an interface, a laser beam generally drifts and tilts away from the path predicted by ray optics, an intriguing consequence of its finite transverse extent. 
Such beam shifts manifest more dramatically for structured light fields, and in particular for optical vortices. 
Upon reflection, a field containing a high-order optical vortex is expected to experience not only geometrical shifts, but an additional splitting of its high-order vortex into a constellation of unit-charge vortices, a phenomenon known as topological aberration.
In this article, we report on the first direct observation of the topological aberration effect, measured through the transformation of a vortex constellation upon reflection. 
We develop a general theoretical framework to study topological aberrations in terms of the elementary symmetric polynomials of the coordinates of a vortex constellation, a mathematical abstraction which we prove to be the physical quantity of interest.
Using this approach, we are able to verify experimentally the aberration of constellations of up to three vortices reflected from a thin metallic film.
Our work not only deepens the understanding of the reflection of naturally occurring structured light fields such as vortex constellations but also sets forth a potential method for studying the interaction of twisted light fields with matter.
\end{abstract}

\maketitle

\textit{Introduction.\textemdash} The reflection of light beams on interfaces is a quintessential problem in wave optics.
In contrast to simple geometrical optics laws, where a ray of light reflects of a surface with the same incidence angle but on the opposing side of the surface normal, the reflection of beams in wave optics encompasses a more complex behavior. 
Generally, a beam can be mathematically described by a finite spectrum of plane waves. Upon reflection, each of these plane waves individually follows paths determined by geometrical optics \cite{JacksonBook}. However, the changes to the amplitudes and phases of the component plane waves result in a reflected field which is macroscopically different from the incident field. Perhaps the most well-known examples of such phenomena are optical beam shifts, which are spatial and angular deviations of reflected laser beams from their expected ray-optical trajectories \cite{Bliokh2013,Aiello2012}.
Such shifts are usually separated into Goos-Hänchen (GH) \cite{Goos1947,Merano2009} and Imbert-Fedorov (IF) \cite{Bliokh2013,Jayaswa2014} shifts, which occur along and orthogonal to the plane of incidence, respectively. These effects have distinct physical origins: while the GH shifts come from the angular variation of the reflection coefficient of the interface, the IF shifts, also known as the spin-Hall effect of light \cite{Bliokh2015},  stem from the rotation of the plane of incidence experienced by elliptically polarized waves upon reflection. The GH and IF shifts have been extensively studied for different types of optical beams and interfaces in a range of contexts, as discussed in detail in Ref.\cite{Bliokh2013}.

A more subtle topic is the reflection of vortex beams \cite{Bliokh2013,Okuda2006,Okuda2008,Fedoseyev2001,Fedoseyev2008,Bliokh2009,Dasgupta2006}. 
Such beams are most well known for their ability to carry orbital angular momentum (OAM), which is accompanied by a twisted phase structure, i.e., a varying phase front transverse to the propagation direction as $\exp(i\ell\phi)$, where $\ell$ is an integer number and $\phi$ labels the azimuthal angle \cite{Allen1992,RubinszteinRoadmap}. 
The wavefronts of such beams consist of $|\ell|$ identical helicoids nested on the propagation axis, on which lies a strength-$\ell$ optical vortex or phase singularity \cite{Nye1974,Berry2004}. Due to the OAM-induced mixing of the spatial and angular GH and IF shifts \cite{Merano2010}, vortex beams also acquire OAM sidebands upon reflection \cite{Loffler2012} and feature significant deformations in their intensity profiles \cite{Okuda2006,Okuda2008} at critical angles. Beyond beam shifts, Dennis and Götte showed in their seminal work \cite{Dennis2012} that a pure strength-$\ell$ optical vortex splits into a constellation of unit strength vortices by reflecting at a simple dielectric interface, which they recognized as a topological aberration effect. 
It is well known, however, that it is impossible to generate perfect higher-order vortices, as the latter are unstable under any kind of pertubation\cite{Freund1999,Ricci2012,Neo2014}.
Therefore, a fundamental question remains open: how do vortex constellations, the actual observable physical objects, experience topological aberrations?

In this article, we adress this question first by generalizing the 
framework developed in Ref. \cite{Dennis2012} to arbitrary uniform aberrations of vortex constellations, thereby addressing typical experimental limitations in the generation of vortex beams.
We then show that the aberration of a vortex constellation is captured by a linear transformation of the elementary symmetric polynomials of its coordinates, and that this transformation is related to the angular Wirtinger derivatives of the aberration.
Lastly, we detail the experimental observation of the topological aberration of vortex constellations  under total internal reflection from a thin Au film-Fused Silica interface, where aberration effects are enhanced due to the resonant excitation of surface plasmons at the interface. With this method, we are able to verify the topological aberration using constellations with up to $3$ vortices, in good agreement with the theoretical model developed.
Due to the direct link between the vortex dynamics and the properties of the material upon which the light is reflected, our results could be applied to advanced material characterization techniques. 
Moreover, the underlying theoretical description of the vortex dynamics might also be applicable to other fields of physics such as Bose-Einstein condensates \cite{Butts1999,Fetter2009}, superfuilds\cite{Salomaa1987,Steinberg2022}, or topological field theories\cite{Witten1988}.

\textit{Theory of topological aberrations.\textemdash} We start by considering an arbitrary input field containing a constellation of $\ell_m$ identical unit-strength vortices. In momentum space, the scalar part of such a field can be written as \cite{Dennis2012}
\begin{equation}
\tilde{\psi}_I(\boldsymbol{\chi})=\sum_{\ell=0}^{\ell_m} \sigma_\ell(|\chi|)\,\left(\frac{\chi}{|\chi|}\right)^\ell\,,
    \label{EQ_InputFieldRaw}
\end{equation}
where $\boldsymbol{\chi}=(\chi,\chi^*)$, with $\chi=(k_x+ik_y)/\sqrt{2}$, and $(k_x,k_y)$ are Cartesian coordinates in momentum space. $\sqrt{2}|\chi|=k_\perp=\sqrt{k_x^2+k_y^2}$ and $\textrm{Arg}(\chi)=\alpha=\tan^{-1}(k_y/k_x)$, where $k_\perp$ and $\alpha$ are the momentum radial and azimuthal coordinates, respectively. Equation~\eqref{EQ_InputFieldRaw} can be seen as a superposition of optical vortices with topological charges $0\leq\ell\leq \ell_m$ , whose background functions $\sigma_\ell(|\chi|)$ determine both the OAM spectrum and the field's radial features. Furthermore, the constellation coordinates are encoded in the complex roots of the field \eqref{EQ_InputFieldRaw}, which is a polynomial of order $\ell_m$ in $\chi$.

A field of the form given in Eq.~\eqref{EQ_InputFieldRaw}, i.e. a constellation of vortices, is naturally obtained by attempting to generate a vortex of order $\ell_m$ experimentally, due to inherent limitations in light-shaping devices and/or subsequent aberrations caused by mirrors, lenses, and other refractive elements. Nevertheless, for small aberrations, which is the case for a carefully assembled experiment, the contributions of $\ell<\ell_m$ are small and the constellation is tightly confined to the beam's central propagation direction. In this case, we can approximate the field in Eq.~\eqref{EQ_InputFieldRaw} in real space to the lowest order of each OAM component, obtaining
\begin{equation}
\psi_I(\boldsymbol{\xi})=\sum_{\ell=0}^{\ell_m} \bar{\sigma}_\ell \xi^\ell\,,
\label{EQ_InputFieldReal}
\end{equation}
where $\boldsymbol{\xi}=(\xi,\xi^*)$, with $\xi=(x+iy)/\sqrt{2}$, and where $(x,y)$ are the Cartesian coordinates in real space. The coefficients $\bar{\sigma}_\ell$, derived in detail in the Supplementary Material, represent the $\ell$-th order moments of the background functions $\sigma_\ell(|\chi|)$. 

Upon a spatially uniform aberration such as the reflection from a flat interface, each plane wave component of momentum $\boldsymbol{\chi}$ in the angular spectrum \eqref{EQ_InputFieldRaw} is transformed as $\exp(i\boldsymbol{\chi}\cdot\boldsymbol{\xi}^*)\rightarrow\,R(\chi)\exp(i\chi\cdot\xi^*)$, where $R(\boldsymbol{\chi})$ is the momentum-dependent aberration function. In the case of reflection from a flat interface, the aberration function is the reflection coefficient given by the Fresnel coefficients, which also depends on the incident and measurement polarizations \cite{Dennis2012,Bliokh2013}. The resulting field in real space can then be modeled as
\begin{equation}
\begin{aligned}
    \psi_R(\boldsymbol{\xi})=\int d^2\boldsymbol{\chi} \tilde{\psi}_I(\boldsymbol{\chi})R(\boldsymbol{\chi}) \exp(i\boldsymbol{\chi}\cdot \boldsymbol{\xi}^*)\,,
\end{aligned}
\label{EQ_RefFieldRaw}
\end{equation} 
which is a Fourier transform in the complex coordinates $\boldsymbol{\xi}$ and $\boldsymbol{\chi}$. Hence, we can expand the aberration in a power series and use the differentiation property of the Fourier transform to obtain
\begin{eqnarray}
\psi_R(\boldsymbol{\xi})&=&\sum_{n=0}^\infty\sum_{m=0}^n i^{-n} C^m_n \frac{\partial_n \psi_I(\boldsymbol{\xi})}{\partial \xi^{*m}\partial \xi^{n-m}}\,, \label{EQ_RFieldExpRaw}\\
C^m_n &=&\frac{1}{n!} \binom{n}{m} \frac{\partial_n R(\boldsymbol{\chi})}{\partial \chi^{m}\partial \chi^{*n-m}}\Bigg |_{\chi,\chi^*=0}\,,\label{EQ_RExpCoeffRaw}
\end{eqnarray}
where ${\partial}/{\partial \xi}=\left({\partial}/{\partial x}-i{\partial}/{\partial y}\right)/\sqrt{2}$ and ${\partial}/{\partial \chi}=\left({\partial}/{\partial k_x}-i{\partial}/{\partial k_y}\right)/\sqrt{2}$ are Wirtinger derivatives \cite{OsgoodBook}. From Eqs. \eqref{EQ_InputFieldReal} and \eqref{EQ_RFieldExpRaw}, we arrive at the final expression for the aberrated field
\begin{equation}
\begin{aligned}
\psi_R(\boldsymbol{\xi})&=\sum_{\ell=0}^{\ell_m} \bar{\gamma}_\ell \xi^\ell=\sum_{\ell=0}^{\ell_m} \sum_{n=0}^\ell i^{-n}\bar{\sigma}_\ell n!\binom{\ell}{n} C^0_n \xi^{\ell-n}\,,
\end{aligned}
\label{EQ_RefFieldFinal}
\end{equation}
where $\bar{\gamma}_\ell$ are the coefficients of the vortex expansion of the aberrated field.

\begin{figure}
    \centering
    \includegraphics[width=1.04\linewidth]{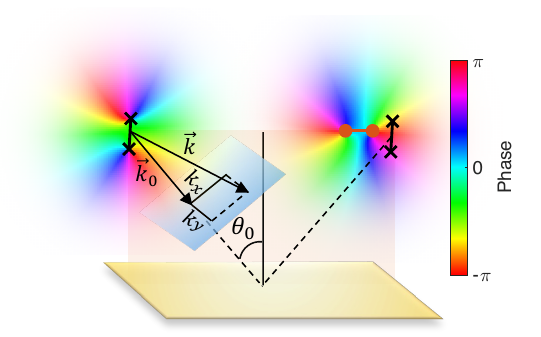}
    \caption{Conceptual picture of the topological aberration of a vortex constellation upon reflection. $\vec{k}$ denotes a plane wave component of the incident vortex, while $\vec{k}_0$ is the central wave vector in the beam. The plane of incidence is $k_xk_z$, and the dashed lines show the ray optics trajectory of the central plane wave $\vec{k}_0$, with the angle of incidence $\theta_0$. The input and reflected constellations are marked with black crosses and orange circles, respectively.}
    \label{Fig_concept}
\end{figure}

Equation \eqref{EQ_RefFieldFinal} shows that an aberration decomposes the incoming light field into a superposition of its Wirtinger derivatives, weighted by the Wirtinger derivatives of the aberration function. Interestingly, in the case of the vortex input field of Eq.\eqref{EQ_InputFieldRaw} the derivative modes are simply optical vortices of lower order, in such a way that the aberrated field still contains a collection of $\ell_m$ phase singularities, but in a deformed constellation, as we illustrate in Fig.~\ref{Fig_concept}. This implies that by monitoring the changes in a vortex constellation, one can gain insight into the properties of the aberration function, and thus into features of the light-matter interaction causing the aberration.   

\textit{High-order aberrations of vortex constellations\textemdash} Equation \eqref{EQ_RefFieldFinal} establishes a direct correspondence between the coefficients of the vortex expansions of the incident and aberrated fields. However, the question still remains as to how the coordinates of the corresponding constellations compare. This problem is elegantly solved by Vieta's theorem \cite{Cohn2000Algebra}, which we illustrate in the following. Consider an arbitrary constellation of $\ell_m$ vortices with coordinates $(x_j,y_j)$, which yields a vortex expansion with roots $\Delta_j=(x_j+iy_j)/\sqrt{2}$ and coefficients $c_i$ to be determined. According to Vieta's theorem we have that 
\begin{equation}
    e_j=(-1)^j\frac{{c}_{\ell_m-j}}{{c}_{\ell_m}}\,, \quad 1<j\leq\ell_m\,,
    \label{EQ_SymPol}
\end{equation}
where $e_j$ is the $j$-th order elementary symmetric polynomial (ESP) of the set of complex roots $\{\Delta\}$. In terms of ESPs, equation \eqref{EQ_RefFieldFinal} assumes the simple form
\begin{equation}
    \boldsymbol{e}_R=\hat{R}_\chi\boldsymbol{e}_I\,,
    \label{EQ_JordanFinal}
\end{equation}
where the vectors $\boldsymbol{e}_I$ and $\boldsymbol{e}_R$ contain the ESPs of the input and aberrated constellations, respectively, with the aberration operator $\hat{R}_\chi$ acting on the $\ell_m$-dimensional subspace spanned by the ESPs. The explicit form for the operator $\hat{R}_\chi$ is an upper triangular matrix containing the Wirtinger derivatives of the aberration function $R(\boldsymbol{\chi})$ up to the order $\ell_m$, as we detail in the Supplementary Material.

We conclude from Eq.~\eqref{EQ_JordanFinal} that the aberration of a vortex constellation is fully captured by a linear transformation of its ESPs. Furthermore, an intuitive physical/geometrical interpretation of the ESP transformations is possible by means of Newton's identities \cite{Cohn2000Algebra}, which relate the ESPs to power sums of a set of complex roots. For example, $e_1$ is proportional to the constellation barycenter, whose transformation contains the GH and IF beam shifts of vortex beams \cite{Bliokh2013,Aiello2012,Dennis2012,Merano2010}. On the other hand, $e_2$ and $e_3$ give the second and third moments of the constellation positions with respect to the barycenter, which are related to the variance and the skewness of the constellation, respectively. For higher orders, the ESPs are sums of products of moments of the constellation positions that add up to that order, whose meaning we could not identify. 

For clarity, let us explicitly consider the case of $\ell_m=2$. We choose the origin of the reference frame at the barycenter of the input constellation, such that the input field is $\psi_I(\boldsymbol{\xi})\propto\xi^2+e_{I2}\xi^0$. From Eqs.~\eqref{EQ_RefFieldFinal} and \eqref{EQ_SymPol}, we obtain that
\begin{eqnarray}
    e_{R1}&=&2 i R_\chi^\prime/R_\chi\Big|_0 \,, \label{EQ_e1}\\
    e_{R2}&=&e_{I2}-R_\chi^{\prime\prime}/R_\chi\Big|_0\,,\label{EQ_e2} 
\end{eqnarray}
where $R_\chi^{\prime}|_0$ and $R_\chi^{\prime \prime}|_0$ are the first and second Wirtinger derivatives of $R(\boldsymbol{\chi})$ at $\chi=\chi^*=0$. Equation \eqref{EQ_e1} shows the shift of the barycenter of the constellation by the known Artmann translator of a unit vortex \cite{Merano2010,Bliokh2009}, which is the first order topological aberration of the constellation. In fact, this result, previously predicted only for perfect input vortices \cite{Dennis2012}, holds regardless of the particular geometry or the order of a vortex constellation. Furthermore, equation~\eqref{EQ_e2}, shows the second-order topological aberration, which amounts to the stretching of the input constellation and its rotation around the barycenter. 
\\

\begin{figure}
    \centering
    \includegraphics[width=1\linewidth]{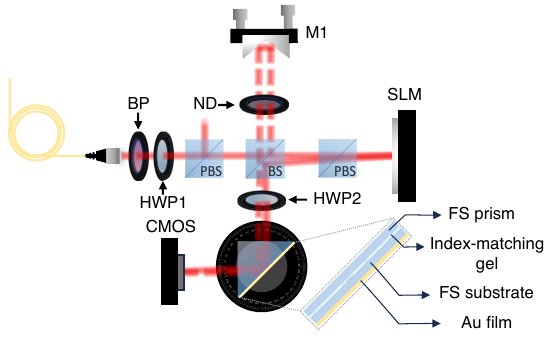}
    \caption{Experimental setup containing: BP, bandpass filter; ND, neutral density filter; M1, retroreflecting mirror; PBS, polarizing beam-splitter; BS, 50:50 non-polarizing beam-splitter; HWP1 and HWP2, true-zero order half-wave plates; SLM, spatial light modulator; CMOS, camera sensor. The inset shows a breakdown of our reflecting object, which consists of an Au film assembled on a FS substrate and optically contacted to a FS prism. }
    \label{Fig_ExpSetup}
\end{figure}

\textit{Experiment.\textemdash} To observe the aberration of vortex constellations, we use the experimental setup depicted in Fig.\ref{Fig_ExpSetup}. We use a fiber-coupled diode laser centered at the wavelength $\lambda= 810\,$nm, operated below the lasing threshold and filtered by a narrow bandpass filter (Semrock, 3nm bandwidth). The beam is then polarized on a polarizing beam-splitter (PBS) and divided with a non-polarizing  50:50 beam-splitter (BS) into a probe beam, which we prepare with the desired constellation, and a reference beam, used afterwards for the generation of off-axis holograms. To prepare the probe beam, we shape it on a spatial light modulator (SLM, Holoeye Pluto 2.0) displaying a $\exp (i\ell_m \phi)$ phase structure, which naturally leads to a constellation tightly confined to the beam center. We further tune the constellation by introducing wavefront aberrations by means of Zernike polynomials \cite{Lakshminarayanan2011}, ensuring that all the $\ell_m$ vortices are clearly observable in our apparatus. As an additional precaution, we use a second PBS to prevent any undesired changes in polarization due to the SLM.

Our aberrating device consists of a 35nm thick Au film assembled on the hypothenuse of a fused silica (FS) right-angle prism, in the so-called Kretschmann-Raether (KR) configuration \cite{KretschmannRaether,Akimov2017}. The Au film is deposited on an FS substrate using electron beam-assisted evaporation and optically contacted to the prism with index-matching gel (Thorlabs G608N3). In the KR configuration, a P-polarized input beam can resonantly excite surface plasmons at the interface between FS and the Au film via attenuated total reflection (ATR), enhancing the angular gradients of the reflection coefficient, and hence the topological aberration imprinted on the reflected beam \cite{Yin2004}. By measuring the power of the reflected S- and P-polarized probe beams as a function of the angle of incidence and using a transfer matrix model (see Supplementary Material) to numerically fit the results, we determine the resonant ATR angle and the effective permittivity of the Au film to be $\theta_{ATR}\approx45.02^\circ$ and $\epsilon_f=-22.8456+1.2619 i$, respectively. Furthermore, since only P-polarized light excites surface waves, we can use the reflected S-polarized field as a good approximation of the input field and consider the aberration function $R(\boldsymbol{\chi})$ to be the fitted reflection coefficient for P-polarized light. 

\begin{figure*}
    \centering    \includegraphics[width=\linewidth]{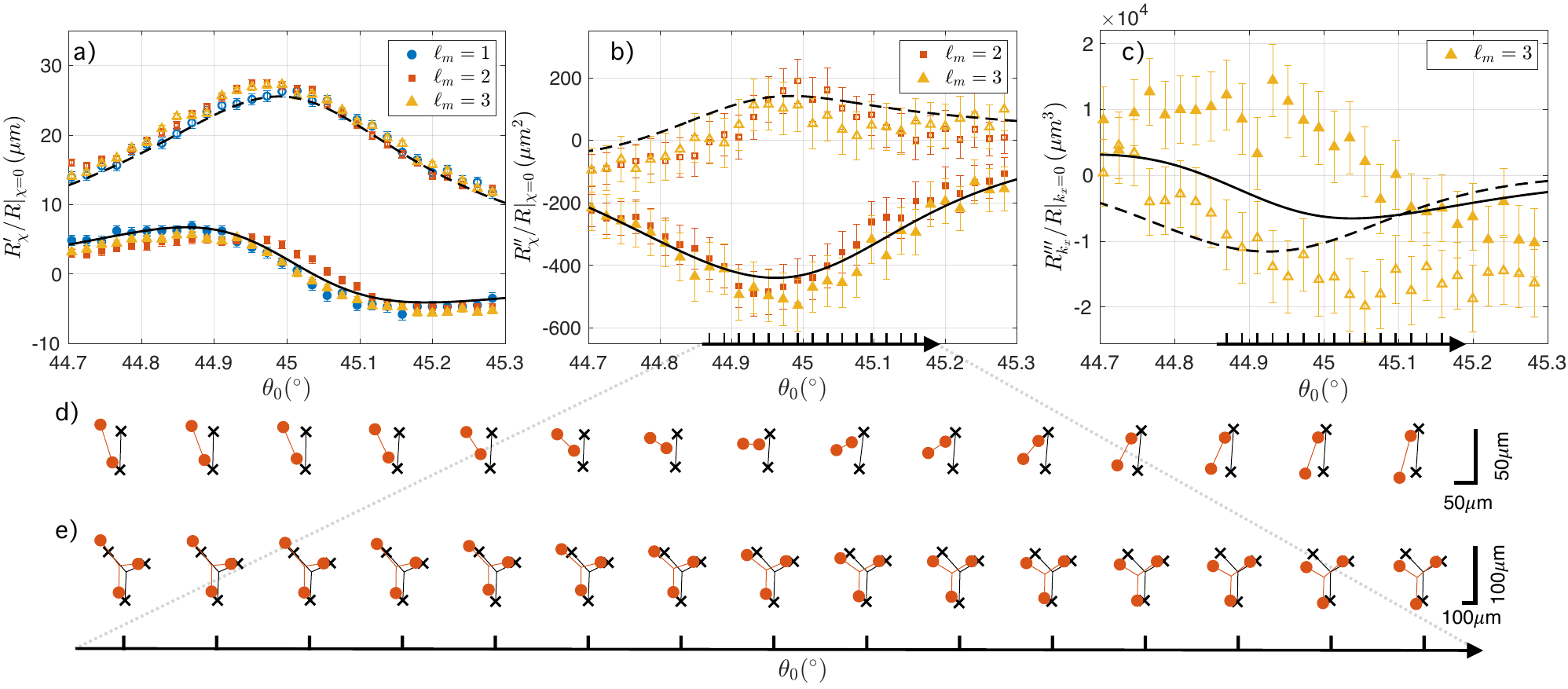}
    \caption{a) First, b) second, and c) third Wirtinger derivatives of the reflection coefficient for P-polarized light, retrieved from the measured vortex constellations for $1\leq \ell_m \leq 3$. In each plot, the real (imaginary) part is shown as filled (hollow) markers for measured data, and as solid (dashed) curves for theoretical simulations. Additionally, in d) and e) we show the measured vortex constellations of the input field (crosses) and the reflected field (circles) with $\ell_m = 2$ in d) and $\ell_m=3$ in e). The measured constellations are shown for the 15 angles of incidence highlighted on the $x$ axis of b) and c).}
    \label{Fig_ExpResults}
\end{figure*}

To measure the constellation coordinates, we superpose the probe and reference beams off-axis and record the interference patterns with a CMOS camera (IDS xxx, 2.2 $\mu$m pixel size), from which both the intensity and phase profiles are digitally retrieved \cite{Goodman1967,Verrier2011}. It is worth noting that the measured constellations are affected by interference with stray light coming from the protective glass window on the CMOS sensor, the reflective layers of the beam-splitters, etc., which cannot be eliminated in post-processing due to the propagation direction being almost identical to the probe field. The laser was operated below the threshold in order to eliminate these unwanted interference effects, at the cost of requiring precise control of the path length difference between reference and probe beams. In our setup we match the path lengths with a translation stage, also ensuring negligible lensing effect on the recorded holograms. Furthermore, we dim the intensity of the reference beam with neutral density filters, improving the hologram visibility around the singularity positions and allowing better use of the CMOS dynamic range. Finally, we determine the constellation coordinates from the retrieved phase profiles using the algorithm detailed in the Supplementary Material. The uncertainty of the method was estimated to be $1/2$ pixel irrespective of the hologram fringe density, which is the uncertainty we used to calculate the error bars shown in Fig.\ref{Fig_ExpResults}.

\textit{Results.\textemdash} We show in Fig \ref{Fig_ExpResults} our experimental results for constellations with $1\leq\ell_m\leq 3$. In Fig.\ref{Fig_ExpResults}a we show the logarithmic Wirtinger derivative of the reflection coefficient, retrieved from the shifts in the barycenters of the constellations using Eqs.~\eqref{EQ_RefFieldFinal} and \eqref{EQ_e1}. We see that the retrieved derivatives accurately follow the theoretical curves simulated from the measured material parameters under the assumption $R_{\chi}^{(n)}|_0\approx R_{k_x}^{(n)}|_0/2^{n/2}$. Such an approximation is reasonable in our case, since $k_xk_z$ is the plane of incidence, and the constellations are prepared and measured in the same polarization. We note that, unlike the shifts in the center of mass for the vortex-carrying fields \cite{Dennis2012,Bliokh2009}, the measured barycenter shifts do not depend on the total topological charge $\ell_m$. This result is a direct measurement of the first-order topological aberration of a vortex constellation, which, despite its strong correspondence to the GH and IF shifts, had not yet been observed. Furthermore, the barycenter shifts we report are greatly enhanced compared to those of dielectric interfaces, reaching more than $30$ wavelengths in magnitude. A similar enhancement of the GH shift has been reported in \cite{Yin2004} for Gaussian beams, which here we extend to vortex beams and out-of-plane (IF) shifts. 

Coming to the second-order topological aberration, in Fig.~\ref{Fig_ExpResults}b we show the second Wirtinger derivative of the reflection coefficient, retrieved from the transformations in the second-order ESPs of the constellations with $\ell_m=2$ and $\ell_m=3$. The retrieved derivatives agree well with the theoretical simulations, although the experimental results show sharper features when compared to the theory. The geometrical meaning of the second-order topological aberration can also be seen in Fig.~\ref{Fig_ExpResults}d, where we show the measured input and aberrated constellations for the $15$ angles of incidence highlighted by an arrow in Fig.~\ref{Fig_ExpResults}b. We see that near the resonant ATR angle, both the size and shape of the constellation change noticeably, in addition to the barycenter shifts following Fig.~\ref{Fig_ExpResults}a. Furthermore, we note that the reflected S-polarized constellations do not change noticeably with the angle of incidence, supporting our assumption that they are good approximations of the input constellations.

Lastly, in Fig.~\ref{Fig_ExpResults}c we show the third Wirtinger derivative of the reflection coefficient, retrieved from the changes in the third-order ESPs of the $\ell_m=3$ constellation. In this case, the experimental results diverge more substantially from the theoretical simulations, which we attribute to two possible sources of error. First, measuring constellations is increasingly challenging as the constellation order increases, since the intensity near the singularity positions decreases exponentially with $\ell_m$. For $\ell_m>2$, in our case, properly exposing the hologram near the singularity positions requires long exposure times, and hence the measurements are more susceptible to mechanical instabilities and noise. Second, the transfer matrix model that we used for the theoretical simulations does not include the roughness of the Au film surface. Due to these limitations, we did not measure the aberrations of constellations of even higher order. Nonetheless, analogously to before, we show in Fig.~\ref{Fig_ExpResults}e the raw constellation coordinates for the set of angles highlighted by an arrow in Fig.~\ref{Fig_ExpResults}c. Here, the deformation of the P-polarized constellations as well as the shift of the barycenter can be clearly seen.

\textit{Conclusion.\textemdash}  In this article, we developed a new framework to describe the topological aberration of vortex constellations and reported on the first direct observation of the phenomenon predicted more than a decade ago \cite{Dennis2012}. We showed that aberrations generally affect the elementary symmetric polynomials of a constellation's coordinates, from which the angular Wirtinger derivatives of the aberration can be directly retrieved. We demonstrate the effect experimentally by measuring the topological aberrations of vortex constellations with up to $3$ vortices upon attenuated total reflection, where aberration effects are enhanced by the resonant excitation of surface plasmon polaritons on the interface between a thin metallic film and a dielectric medium. Our results mark the first experimental observation of the topological aberration effect and introduce a new framework for probing light-matter interactions with twisted light.

These topological aberration effects, when described with a precise theoretical model, will be a useful tool to derive the optical properties of metal-dielectric interfaces, thereby possibly simplifying current standard characterization techniques such as ellipsometry.
In addition, because of the generality of our underlying theoretical framework describing the dynamics of vortex constellations, it will be interesting to apply our approach to other singular light fields featuring, for example, polarization singularities \cite{Berry2004}, vortex knots \cite{Leach2004} and polarization knots \cite{Larocque2018}.
Beyond light waves, we expect our work to inspire connections to other fields of physics where complex vortex constellations appear, such Bose-Einstein condensates, superfluids \cite{Butts1999,Fetter2009, Salomaa1987, Steinberg2022}, and even topological field theories \cite{Witten1988}.
Finally, as the structured light fields are seen as promising approaches to encode classical and quantum information, a better understanding / simplified description of the vortex dynamics will facilitate the reduction of errors caused by aberrations \cite{al2021structured}.

\textit{Acknowledgements.\textemdash}We thank Jörg Gotte and Mark Dennis for inspiring
discussions at the ICOAM 2022. We also thank Matias Eriksson for the helpful comments on the data processing. R. B. acknowledges the support of the Academy of Finland through the postdoctoral researcher funding (decision 349120). M. H. acknowledges support from the Doctoral School of Tampere University, Emil Aaltonen foundation, and the Magnus Ehrnrooth foundation through its graduate student scholarship R. F.  acknowledges the support of the Academy of Finland through the Academy Research Fellowship (decision 332399). All authors acknowledge the support of the Academy of Finland through the Competitive Funding to Strengthen University Research Profiles (decision 301820) and the support of the Photonics Research and Innovation Flagship (PREIN - decision 320165).
\

\bibliography{references.bib}

\end{document}


\preprint{APS/123-QED}

\title{Supplementary material for the article ``High-order aberrations of vortex constellations"}

\author{R. F. Barros}
\affiliation{Tampere University, Photonics Laboratory, Physics Unit, Tampere, FI-33720, Finland}
\author{S. Bej}%
\affiliation{Tampere University, Photonics Laboratory, Physics Unit, Tampere, FI-33720, Finland}
\author{M. Hiekkamäki}%
\affiliation{Tampere University, Photonics Laboratory, Physics Unit, Tampere, FI-33720, Finland}
\author{M. Ornigotti}%
\affiliation{Tampere University, Photonics Laboratory, Physics Unit, Tampere, FI-33720, Finland}
\author{R. Fickler}
\affiliation{Tampere University, Photonics Laboratory, Physics Unit, Tampere, FI-33720, Finland}

\date{\today}
\maketitle

\section{Derivation of the vortex expansion of input and reflected fields}
\subsection{Input field}

The input field described in the main text is  the following 
\begin{equation}
\tilde{\psi}_I(\boldsymbol{\chi})=\sum_{\ell=0}^{\ell_m} \sigma_\ell(|\chi|)\,\left(\frac{\chi}{|\chi|}\right)^\ell\,,
    \label{EQ_InputFieldRaw}
\end{equation}
where $\boldsymbol{\chi}=(\chi,\chi^*)$, with $\chi=(k_x+ik_y)/\sqrt{2}$, and where $(k_x,k_y)$ are Cartesian coordinates in momentum space. Note that $\sqrt{2}|\chi|=k_\perp=\sqrt{k_x^2+k_y^2}$ and $\textrm{Arg}(\chi)=\alpha=\tan^{-1}(k_y/k_x)$, where $k_\perp$ and $\alpha$ are the radial and azimuthal coordinates in the momentum space, respectively. To obtain the field in real space we simply take the Fourier transform of \eqref{EQ_InputFieldRaw}, obtaining 
\begin{equation}
\begin{aligned}
    \psi_I(\boldsymbol{\xi})&=\int d^2\boldsymbol{\chi} \tilde{\psi}_I(\boldsymbol{\chi}) \exp(i\boldsymbol{\chi}\cdot \boldsymbol{\xi}^*)\,,\\
    &=\int_0^\infty dk_\perp\int_0^{2\pi} d\alpha \,k_\perp \tilde{\psi}_I(\boldsymbol{\chi}) \exp(i\boldsymbol{\chi}\cdot \boldsymbol{\xi}^*)\,,\\
    &=\sum_{\ell=0}^{\ell_m}\int_0^\infty dk_\perp\,k_\perp \sigma_\ell(k_\perp)\left(\int_0^{2\pi} d\alpha e^{i[\ell\alpha+2k_\perp r_\perp\cos(\alpha-\phi)]} \right)\,,    
\end{aligned}
\label{EQ_InputFieldReal}
\end{equation}
where we used the relation $d\chi d\chi^*=dk_xdk_y=k_\perp dk_\perp d\alpha$ derived from the Jacobian of the transformation. In the above equation we have $\boldsymbol{\xi}=(\xi,\xi^*)$, where $\xi=(x+iy)/\sqrt{2}$, and $(x,y)$ are the Cartesian coordinates in the real space, as used in the main text. As before, we have $\sqrt{2}|\xi|=r_\perp=\sqrt{x^2+y^2)}$ and $\textrm{Arg}(\xi)=\phi=\tan^{-1}(y/x)$, where $r_\perp$ and $\phi$ are the radial and azimuthal coordinates in real space, respectively. 

The azimuthal integral in \eqref{EQ_InputFieldReal} results in
\begin{equation}
    \int_0^{2\pi} d\alpha e^{i[\ell\alpha+2k_\perp r_\perp\cos(\alpha-\phi)]}=2\pi i^\ell J_\ell(k_\perp r_\perp )e^{i\ell\phi}\,,
\end{equation}
\label{EQ_AzimuthInt}
where
\begin{equation}
J_\ell(k_\perp r_\perp )= \sum_{m=0}^\infty \frac{(-1)^m}{m!(m+\ell)!}\left(\frac{k_\perp r_\perp }{2} \right)^{2m+\ell}\,,
    \label{EQ_Bessel}
\end{equation}
is the $\ell$-th order Bessel function of the first kind. The key here is to notice that the lowest power of $k_\perp r_\perp$ in $J_\ell(k_\perp r_\perp )$ is $\ell$. Using this lowest-order approximation, Eq.\eqref{EQ_InputFieldReal} becomes 
\begin{equation}
\psi_I(\boldsymbol{\xi})=\sum_{\ell=0}^{\ell_m} \bar{\sigma}_\ell \xi^\ell\,,
\label{EQ_InputFieldFinal}
\end{equation}
where
\begin{equation}
        \bar{\sigma}_\ell=\frac{i^\ell}{\ell !}\int d^2\boldsymbol{\chi} \,\sigma_\ell(|\chi|) |\chi|^\ell\,.\\
        \label{EQ_Sigma}
\end{equation}
From the above equations, we conclude that the effect of the lowest-order approximation of the Bessel function \eqref{EQ_Bessel} is to approximate the background fields as $\sigma_\ell(|\chi|)~\propto~\bar{\sigma}_\ell|\chi|^\ell$, where $\sigma_\ell$ is proportional to the overlap between the true and approximated background fields. 

\subsection{Reflected field}

The reflected field is obtained by multiplying each plane wave component of the angular spectrum \eqref{EQ_InputFieldRaw} of the input field by a momentum-dependent aberration function $R(\boldsymbol{\chi})$, from which we obtain
\begin{equation}
\begin{aligned}
    \psi_R(\boldsymbol{\xi})=\int d^2\boldsymbol{\chi} \tilde{\psi}_I(\boldsymbol{\chi})R(\boldsymbol{\chi}) \exp(i\boldsymbol{\chi}\cdot \boldsymbol{\xi}^*)\,,
\end{aligned}
\label{EQ_RefFieldRaw}
\end{equation}
which is a Fourier transform in the variables $\boldsymbol{\chi}$ and $\boldsymbol{\xi}$. We can then expand $R(\boldsymbol{\chi})$ in a taylor series of $\chi$ and $\chi^*$ around the origin($\chi=\chi^*=0$), obtaining
\begin{eqnarray}
    R(\boldsymbol{\chi})&=&\sum_ {n=0}^\infty \sum_{m=0}^n C^m_n \chi^m \chi^{*n-m}\,,\label{Eq_RefCoeffExp}\\
C^m_n &=&\frac{1}{n!} \binom{n}{m} \frac{\partial_n R(\boldsymbol{\chi})}{\partial \chi^{m}\partial \chi^{*n-m}}\Bigg |_{\chi,\chi^*=0}\,,\label{EQ_RExpCoeffRaw}
\end{eqnarray}
where
\begin{equation}
\begin{aligned}
    \frac{\partial}{\partial\chi}&=\frac{1}{\sqrt{2}}\left(\frac{\partial}{\partial k_x}  -i\frac{\partial}{\partial k_y} \right)\,,\\
    \frac{\partial}{\partial\chi^*}&=\frac{1}{\sqrt{2}}\left(\frac{\partial}{\partial k_x}  +i\frac{\partial}{\partial k_y} \right)\,,
    \end{aligned}
\end{equation}
are the Wirtinger derivatives in $k$-space. Inserting \eqref{Eq_RefCoeffExp} in \eqref{EQ_RefFieldRaw}, and using the differentiation property of the Fourier transform, we finally obtain    
\begin{equation}
\begin{aligned}
    \psi_R(\boldsymbol{\xi})&=\sum_{n=0}^\infty \sum_{m=0}^n C^m_n\int d^2\boldsymbol{\chi} \tilde{\psi}_I(\boldsymbol{\chi})\chi^m \chi^{*n-m} \exp(i\boldsymbol{\chi}\cdot \boldsymbol{\xi}^*)\,,\\
    &=\sum_{n=0}^\infty \sum_{m=0}^n i^{-n}C^m_n \frac{\partial_n \psi_I(\boldsymbol{\xi})}{\partial \xi^{*m}\partial \xi^{n-m}}\,,\\
    &=\sum_{\ell=0}^{\ell_m} \sum_{n=0}^\ell i^{-n}\bar{\sigma}_\ell n!\binom{\ell}{n} C^0_n \xi^{\ell-n}\,,
\end{aligned}
\label{EQ_RefFieldSteps}
\end{equation}
which is the expression we used in the main text. For a better understanding of the phenomenon, it is beneficial to separate Eq.~\eqref{EQ_RefFieldSteps} in two terms
\begin{equation}
\psi_R(\boldsymbol{\xi})=R(0)\psi_I(\boldsymbol{\xi})+\sum_{\ell=0}^{\ell_m} \sum_{n=1}^\ell i^{-n}\bar{\sigma}_\ell n!\binom{\ell}{n} C^0_n \xi^{\ell-n}\,
\label{EQ_RefFieldFinal}
\end{equation}
where we see that the first term gives precisely what one would expect in the absence of any topological aberration, i.e., the whole beam experiencing the reflection coefficient of its central wave vector. On the other hand, the second term shows the topological aberration effect as a change in the vortex expansion of the input field according to the momentum Wirtinger derivatives of the aberration function.

\subsection{Elementary symmetric polynomials and the connection between input and reflected fields}

To unveil the simple connection between the input and reflected fields, we insert Eqs.~\eqref{EQ_Sigma} and \eqref{EQ_RExpCoeffRaw} into Eq.~\eqref{EQ_RefFieldSteps}, obtaining the following expression
\begin{equation}
    \psi_R(\boldsymbol{\xi})=\sum_{\ell=0}^{\ell_m} \sum_{n=0}^\ell \frac{i^{\ell-n}}{(\ell-n)!}\frac{R_n(0)}{n!}\frac{\bar{\sigma}_\ell\,\ell!}{i^\ell} \xi^{\ell-n}=\sum_{\ell=0}^{\ell_m} \bar{\gamma}_\ell \xi^\ell\,,
    \label{EQ_ConnectionRaw}
\end{equation}
where we used the shortened notation $R_n(0)=\partial^n R(\boldsymbol{\chi})/\partial \chi^n \big|_{\chi=\chi^*=0}$. Our goal is to determine the relationship between the coefficients $\bar{\gamma}_\ell$ of the vortex expansion of the reflected field and the coefficients $\bar{\sigma}_\ell$. For that, we first define the vectors of expansion coefficients $\boldsymbol{\gamma}=(\bar{\gamma}_{\ell_m},\bar{\gamma}_{\ell_m-1},\cdots,\bar{\gamma}_{0})^T$ and $\boldsymbol{\sigma}=(\bar{\sigma}_{\ell_m},\bar{\sigma}_{\ell_m-1},\cdots,\bar{\sigma}_{0})^T$, in terms of which Eq.~\eqref{EQ_ConnectionRaw} can be rewritten as 
\begin{equation}
    \begin{aligned}
    \boldsymbol{\gamma}&=\mathcal{T}R_\chi(\mathcal{J}_{0})\mathcal{T}^{-1}\boldsymbol{\sigma}\,,
    \end{aligned}
    \label{EQ_JordanRaw}
\end{equation}
where $\mathcal{T}_{mn}=i^m/m!\delta_{mn}$, and where
\begin{equation}
    R_\chi(\mathcal{J}_{0})=
    \begin{pmatrix}
    \frac{R_0(0)}{0!}&\frac{R_1(0)}{1!}&\frac{R_2(0)}{2!}&\cdots&\frac{R_{\ell_m}(0)}{\ell_m!} \\
        0&\frac{R_0(0)}{0!}&\frac{R_1(0)}{1!}&\cdots&\frac{R_{\ell_m-1}(0)}{(\ell_m-1)!} \\
                0&0&\frac{R_0(0)}{0!}&\cdots&\frac{R_{\ell_m-2}(0)}{(\ell_m-2)!} \\
                \vdots&\vdots&\ddots&\ddots&\vdots\\
                0&0&0&0&\frac{R_0(0)}{0!}
    \end{pmatrix}\,,
    \label{EQ_RefCoeffJordan}
\end{equation}
is nothing but the function $R_{\chi}(\chi)=\sum_{m=0}^\infty R_{m}(0)\chi^m/m!$ evaluated at the Jordan block matrix $\mathcal{J}_{0mn}=\delta_{m(n+1)}$ \cite{Cohn2000Algebra}. 
Furthermore, by noticing that $\mathcal{J}_0$ is the Jordan canonical form of the matrix $\Lambda=\mathcal{T}\mathcal{J}_0\mathcal{T}^{-1}$, we can simplify Eq.~\eqref{EQ_JordanRaw} to the final form
\begin{equation}
    \boldsymbol{\gamma}=\hat{R}_\chi\boldsymbol{\sigma}\,,
    \label{EQ_JordanFinal}
\end{equation}
where $\hat{R}_\chi=R_\chi(\Lambda)$, derived from the property $\mathcal{T}R_\chi(\mathcal{J}_0)\mathcal{T}^{-1}=R_\chi(\mathcal{T}\mathcal{J}_0\mathcal{T}^{-1})$ \cite{Cohn2000Algebra}.
Equation \eqref{EQ_JordanFinal} shows that the aberration of a field is equivalent to an operator $\hat{R}_{\boldsymbol{\chi}}$ acting on an element $\boldsymbol{\sigma}$ of the finite-dimensional vector space spanned by the vortices $\xi^\ell$ ($0\leq\ell\leq \ell_m$), whose matrix representation is given by Eq. (15) above.

Lastly, we can rewrite Eq.~\eqref{EQ_JordanFinal} in terms of the elementary symmetric polynomials (ESP) introduced in the main text. An arbitrary field of the form $\psi=\sum_{j=0}^{\ell_m}c_j\boldsymbol{\xi}^j$ has a set of $\ell_m$ roots $\Delta_j=(x_j+iy_j)/\sqrt{2}$ $(1\leq j \leq \ell_m)$, each of which represents a coordinate $(x_j,y_j)$ of the vortex constellation. The elementary symmetric polynomials $e_j$ of the set of roots $\Delta_j$ are defined as
\begin{equation}
    e_j=\!\!\!\!\!\!\!\!\!\sum_{1\leq n_1 \leq n_2\cdots \leq n_j\leq \ell_m} \Delta_{n_1}\Delta_{n_2}\cdots \Delta_{n_j}\,, \qquad 1\leq j\leq \ell_m\,,
\end{equation}
from which the complex coefficients $c_j$ can be calculated using the following expression \cite{Cohn2000Algebra}
\begin{equation}
    e_j=(-1)^j\frac{{c}_{\ell_m-j}}{{c}_{\ell_m}}\,, \quad 1<j\leq\ell_m\,,
    \label{EQ_SymPol}
\end{equation}
known as Vieta's formula. Using the ESPs, Eq.~\eqref{EQ_JordanFinal} assumes the form presented in the main text, namely, 
\begin{equation}
    \boldsymbol{e}_R=\hat{R}_\chi\boldsymbol{e}_I\,,
\end{equation}
where $\boldsymbol{e}_{Ij}=(-1)^{j+1}\boldsymbol{\sigma}_j/\sigma_{\ell_m}$ and $\boldsymbol{e}_{Rj}=(-1)^{j+1}\boldsymbol{\gamma}_j/\gamma_{\ell_m}$ are vectors containing the ESPs of the input and aberrated constellations,

\section{Newton's identities and the physical/geometrical interpretation of the elementary symmetric polynomials}

An interesting geometrical interpretation of the ESPs can be obtained from Newton's identities \cite{Cohn2000Algebra}, namely,  
\begin{equation}
\frac{i}{\ell_m} e_i=\sum_{j=1}^i (-1)^{j-1}e_{i-j}\mu_j\,,
    \label{EQ_GirardNewton}
\end{equation}
where we define $e_0=0$, and where  
\begin{equation}
\mu_j=\frac{1}{\ell_m}\sum_{k=1}^{\ell_m} \Delta_{k}^j\,,
    \label{EQ_Moment}
\end{equation}
is the $j$-th moment of the singularity constellation. Considering, without loss of generality, that $e_1=\mu_1=0$, which amounts to changing the origin of the reference frame to the barycenter of the constellation, we directly obtain the following identities
\begin{equation}
    \begin{aligned}
        -\frac{2}{\ell_m}e_2&=\mu_2\,\\
        \frac{3}{\ell_m}e_3&=\mu_3\,,
    \end{aligned}
    \label{EQ_MomentsFinal}
\end{equation}
showing that $e_2$ and $e_3$ correspond, respectively, to the second and third moments of the distribution of constellation coordinates, signifying the variance and the skewness of the distribution. However, it should be noted that the roots $\Delta_j$ are complex variables. Consequently, the variance for the complex roots, for example, contains not only the average distance between the vortices and the barycenter, but also their relative azimuthal positions. In general, the $j$-th order ESP is a sum of products of moments whose orders add up to $j$, which lack an obvious interpretation for $j>3$.

\section{Reflectance measurements and fitting of material parameters}

\begin{figure}[b]
    \centering
    \includegraphics[width=0.8\linewidth]{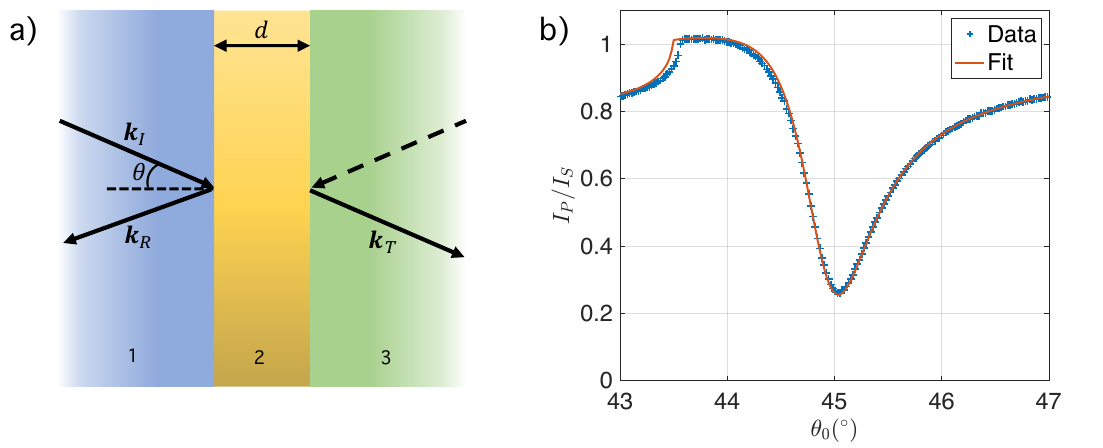}
    \caption{a) Schematic of the transfer matrix formalism for a 3-layer system. The layers 1 and 2 are semi-infinite, while layer 3 has width $d$. The wave vectors $\vec{k}_I$, $\vec{k}_R$ and $\vec{k}_T$ are those of the input, reflected, and transmitted fields, respectively. b) Ratio $I_P/I_S$ between the intensities of P- and S-Polarized light reflected from the Kretschmann-Raether device as a function of the angle of incidence. In the figure, the measured data are plotted with crosses and the least-squares fit using the model \eqref{EQ_RefCoeffModel} is plotted as a solid line. }
    \label{Fig_TMM}
\end{figure}

We model the reflection coefficient of the Kretschmann-Raether (KR) \cite{KretschmannRaether} device used in the experiment with the transfer matrix model \cite{JacksonBook}, as sketched in Fig.\ref{Fig_TMM}a. We consider an input plane wave field $\boldsymbol{E}_{I}=E_I \exp(i \boldsymbol{k}_I\cdot\boldsymbol{r})\boldsymbol{\epsilon}_j$ impinging on layer $2$ from layer $1$ at an angle of incidence $\theta$, with $j=P,S$ denoting the in-plane and out-of-plane polarizations, respectively. Similarly, the reflected and transmitted fields are $\boldsymbol{E}_R=E_R \exp(i \boldsymbol{k}_R\cdot\boldsymbol{r})\boldsymbol{\epsilon}_j$ and $\boldsymbol{E}_T=E_T \exp(i \boldsymbol{k}_T\cdot\boldsymbol{r})\boldsymbol{\epsilon}_j$, respectively. Using the boundary conditions of the electromagnetic field \cite{JacksonBook}, we see that the reflection coefficient is 
\begin{equation}
    R(\theta)=\frac{E_R}{E_I}=\frac{M_{21}}{M_{11}}\,,
\end{equation}
where 
\begin{eqnarray}
    M&=&(M^{(23)}TM^{(12)})^{-1}\,,\\
    M^{(ij)}&=&\frac{1}{2\Lambda_{ij}}
    \begin{pmatrix}
        \Lambda_{ij}+1& \Lambda_{ij}-1 \\
        \Lambda_{ij}-1&\Lambda_{ij}+1
    \end{pmatrix}\,,\\
    \Lambda_{ij}&=&
    \begin{cases}
     \frac{n_j^2/k_z^{(j)}}{n_i^2/k_z^{(i)}}\,, \quad \textrm{P polarization}\\
     \\
     {k_z^{(j)}}/{k_z^{(i)}}\,, \quad \textrm{S polarization}
    \end{cases}
    \\
    T&=&\exp(i\sigma_z k_z^{(2)}d)\,,\\
    k_z^{(j)}&=&k_0\sqrt{n_j^2-\sin(\theta)}
\end{eqnarray}
where $i,j=1,2,3$ label the different layers. Here, $k_0$ is the wave number in vacuum, $n_j$ and $k_z^{(j)}$ are the refractive index and the longitudinal component of the wave vector in layer $j$, respectively, and $\sigma_z$ is the third Pauli matrix. The explicit expression for $R(\theta)$ is 
\begin{equation}
    R(\theta)=\frac{(1-\Lambda_{13})\cos(k_z^{(2)}d)-i(\Lambda_{12}-\Lambda_{23})\sin(k_z^{(2)}d)}{(1+\Lambda_{13})\cos(k_z^{(2)}d)-i(\Lambda_{12}+\Lambda_{23})\sin(k_z^{(2)}d)}\,,
    \label{EQ_RefCoeffModel}
\end{equation}
whose square modulus gives the reflectance of the KR device. 

To obtain the parameters of the Au film used in the experiment, we measured the reflectance of the KR device for both S and P polarizations as a function of $\theta$, using the experimental apparatus described in the main text. We show our result in Fig.~\ref{Fig_TMM}b, along with a least-squares fit of the data using the reflection coefficient modeled by Eq.\eqref{EQ_RefCoeffModel}. For the fitting, we considered the thickness of the Au film to be $d=35nm$, as measured by ellipsometry, and let as free parameters the permittivity of the film and an angular offset, accounting for the backlash of the motorized rotation stage used in the measurements. We obtain the best fit for an effective permittivity $\epsilon_f=-22.8456+1.2619 i$, which differs noticeably from that measured by ellipsometry $\epsilon_{m}=-24.5405 +3.04112 i$.

Lastly, it should be noted that the reflecting surface changes the input polarization vector, and hence the reflection coefficient \eqref{EQ_RefCoeffModel} has corrections depending on the polarizations in which the beam was prepared and measured. In our experiment, however, the light fields are paraxial, and the input and measurement polarizations are jointly set to S or P, in which case the corrections to \eqref{EQ_RefCoeffModel} are negligible. For a more complete discussion on the effects of polarization pre- and post-selection, we encourage the reader to consult Refs.~\cite{Bliokh2013,Dennis2012}.

\begin{figure}[b]
    \centering
    \includegraphics[width=1\linewidth]{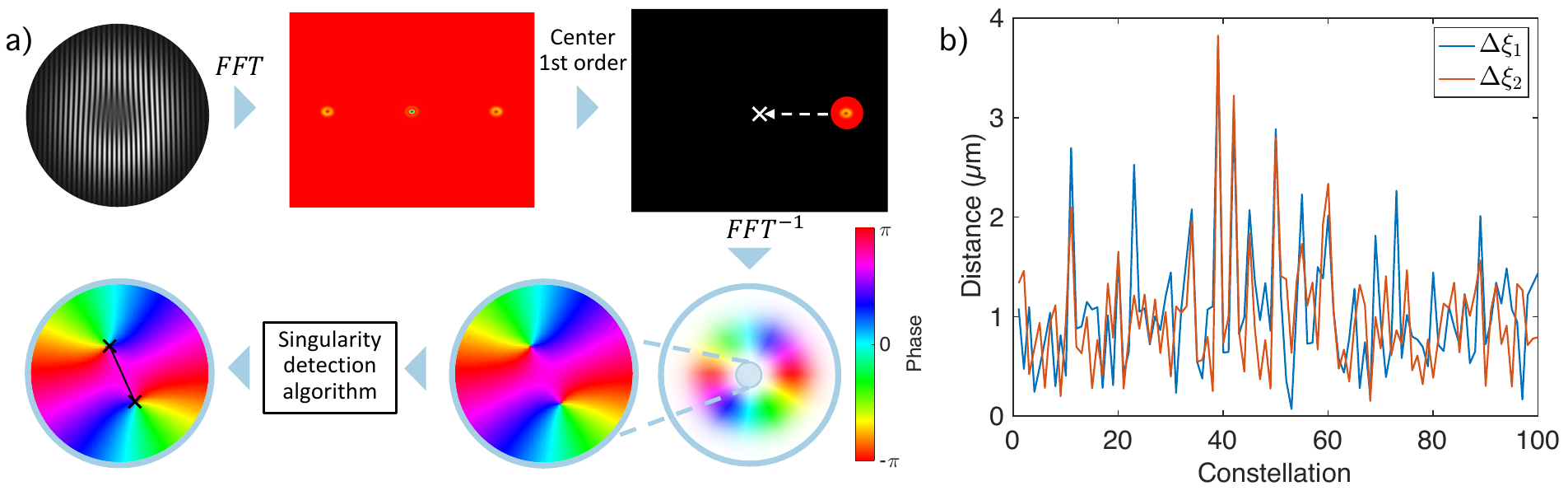}
    \caption{a) Schematics of the image processing of the holograms acquired in the experiment. b) Performance analysis of the singularity detection algorithm. The plot shows the distances $\Delta \xi _1$ and $\Delta \xi _2$ between the measured positions and their true values for 100 random two-point constellations. For each iteration we considered a beam waist of $w_0=320 \mu m$, as in the experiment, with the singularities randomly distributed within a radius of $w_0/10$ from the origin. The simulation grid has a pixel size of $2.2 \mu m$, mimicking the CMOS sensor used in the experiment. }
    \label{Fig_SingDet}
\end{figure}

\section{Image processing and performance analysis of the singularity detection algorithm}

We depict in Fig.\ref{Fig_SingDet}a the workflow we used to process the holograms measured in the experiment. First, we retrieve the phase profile of the probe field from the holograms using off-axis digital holography \cite{Goodman1967,Verrier2011}. The method involves three steps: (1) computing the two-dimensional fast Fourier transform (FFT) of the hologram; (2) selecting the first diffraction order with a binary mask and centering it on the origin of the Fourier space; and (3) performing the inverse FFT. The result of these steps is the approximated complex amplitude distribution of the probe field, provided that the reference field is a collimated Gaussian beam with a beam waist much larger than that of the probe.  

To detect the singularity positions from the retrieved field amplitude, we extract the phase profile and shift it in a 4-pixel closed loop, calculating the phase difference between the shifted image and the previous one at each step. We identify the unit-strength phase singularities in the entire phase profile at once by detecting the pixels around which a phase shift larger than $\pi$ occurred only once. Furthermore, we found that this method is substantially faster than calculating the topological charge around each pixel individually.

We analyze the accuracy of our singularity detection algorithm by processing the holograms of 100 simulated two-point constellations. The positions of the constellations are randomly generated within a radius of $w_0/10$ from the center of the beam, where $w_0=320\mu m$ is the waist of the probe beam. The reference beam is considered as a collimated Gaussian beam with waist $3w_0$ and maximum intensity that matches that of the probe beam, interfering at an angle of $20 mrad$ with the probe. Furthermore, the dimensions of the simulation grid are the same as the CMOS camera used in the experiment, with a pixel size of $2.2\mu m$. 

We show in Fig.\ref{Fig_SingDet}b our simulation results, where we evaluate the distance between the singularity positions measured with our algorithm and the corresponding true values. We obtained an average deviation of $\sigma\approx1\mu m$ for both singularities, which corresponds to $1/2$ of the pixel size. Additionally, we tested our algorithm for different pixel sizes and different angles between the probe and the reference fields (not shown) and observed that the average deviation remains $\approx 1/2$ of the pixel size in the range of the tested parameters.

\bibliography{references.bib}